\documentclass[9pt,twocolumn,twoside]{opticajnl}
\journal{opticajournal} 

\setboolean{shortarticle}{false}

\usepackage{lineno}
\usepackage{graphicx}
\usepackage{subcaption}

\usepackage{physics}
\newcommand{\myvar}{\mathrm{var}}

\usepackage{comment}

\usepackage{xcolor}

\title{Nonlinear squeezing generation via multimode PDC and single photon measurement
}

\author[1,*]{Vojtěch Kala}
\author[2,3]{Denis Kopylov}
\author[1]{Petr Marek}
\author[2]{Polina Sharapova}
\affil[1]{Department of Optics, Palack\'y University, 17. listopadu 1192/12, 77146 Olomouc, Czech Republic}
\affil[2]{Department of Physics, Paderborn University, Warburger Straße 100, D-33098 Paderborn, Germany}
\affil[3]{Institute for Photonic Quantum Systems (PhoQS), Paderborn University, Warburger Straße 100, D-33098 Paderborn, Germany}

 \affil[*]{kala@optics.upol.cz}

\begin{abstract} 
Nonlinear squeezing is a property of non-Gaussian states of light with an important application in continuous variable quantum computing. We study the generation of nonlinear squeezing in multimode systems produced by the photon-added coherent state technique.  We present a protocol and find a regime in which the nonlinear squeezing appears in two modes simultaneously, even for a weak non-Gaussianity induced by the single-photon addition. We explore the properties of nonlinear squeezing depending on the modal structure of light, as well as the seed and local oscillator profiles, and present an optimal measurement strategy.
 
\end{abstract}

\setboolean{displaycopyright}{false} 

\begin{document}

\maketitle

The rich potential of quantum modes of light in generating cluster states has been the subject of extensive study due to its potential applications in quantum computing and quantum information processing~\cite{Larsen,Yokohama2013}. The number of generated modes can be further increased by considering multimode squeezed states~\cite{Kouadou}. 
However, for universal quantum computing, it is necessary to accompany the Gaussian large-scale quantum resources with non-Gaussian elements~\cite{Ra,Chabaud23,Lloyd1999}. At the same time, experimental realization of non-Gaussian resources and operations is quite challenging in quantum optics~\cite{MattiaPRX}. An example of such an operation is a cubic gate~\cite{Braunstein2005}, the lowest order unitary operation that produces nonlinear transformation in the Heisenberg picture. 
In quantum optics, the cubic gate is still not experimentally accessible with sufficient strength, however its measurement-induced version
has been  recently proposed~\cite{Marek2011,Miyata2016} and implemented in the form of nonlinear measurement~\cite{Sakaguchi2023}.

In measurement-induced protocols, the desired operation is usually implemented by using a tailored ancillary state and a suitable measurement accompanied by a feasible  feed-forward~\cite{Zavatta2017,Knight2003,Riabinin2021,Filip2005}. 
For some protocols, such as the measurement-induced squeezing operation~\cite{Filip2005} or the measurement-induced cubic gate operation~\cite{Miyata2016, Sakaguchi2023}, the quality of the ancillary state strongly affects the fidelity of the protocol.
While for the measurement-induced squeezing operation the optimal ancillary state is an ideally squeezed state (the eigenstate of the quadrature operator), for the measurement-induced cubic operation the optimal state is an ideal cubic state, the eigenstate of the operator 
\begin{equation}
\label{eq_nonlop}
\hat{O}(z, \theta) = \hat{P}(\theta)+z\hat{X}^2(\theta),
\end{equation}
where the generalized quadrature operators are $\hat{P}(\theta) = \sin(\theta)\hat{p} + \cos(\theta)\hat{x} $ and $\hat{X}(\theta) = -\cos(\theta)\hat{p} + \sin(\theta)\hat{x} $, while $z$ is a real parameter called cubicity and $\theta$ is an angle of rotation.
The operators $\hat{x}$ and $\hat{p}$ are the quadrature operators and $[\hat{X}(\theta),\hat{P}(\theta)] = [\hat{x},\hat{p}]=i$ holds. 

An ideally squeezed state has infinite squeezing (infinite energy), carries no uncertainty in the given quadrature, and can be Gaussian. Similarly, an ideal cubic state has infinite energy, has no uncertainty in the operator~$\hat{O}$, and is necessarily non-Gaussian.
However, any realistic implementation of either ideally squeezed or ideal cubic state adds the noise to the system that deteriorates the performance of the measurement-induced protocol. 
In case of the cubic operation, once the variance of the used state is above what is achievable by Gaussian states, the protocol can be no longer considered quantum non-Gaussian and thus suitable for quantum computation.

\begin{figure}[ht]
\includegraphics[width=0.49\linewidth]{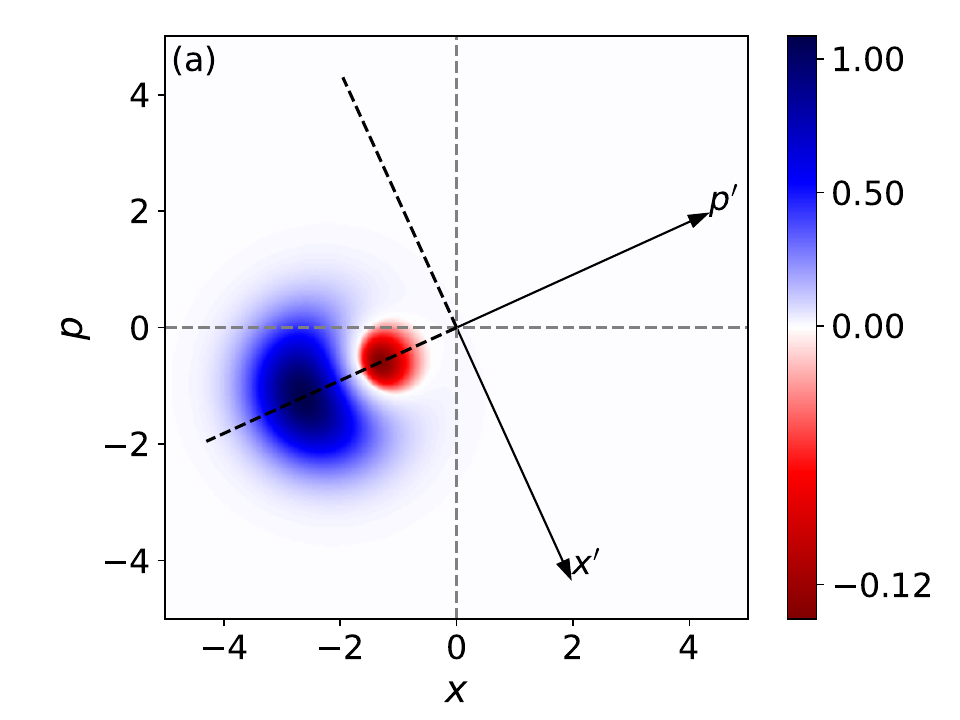}
\includegraphics[width=0.49\linewidth]{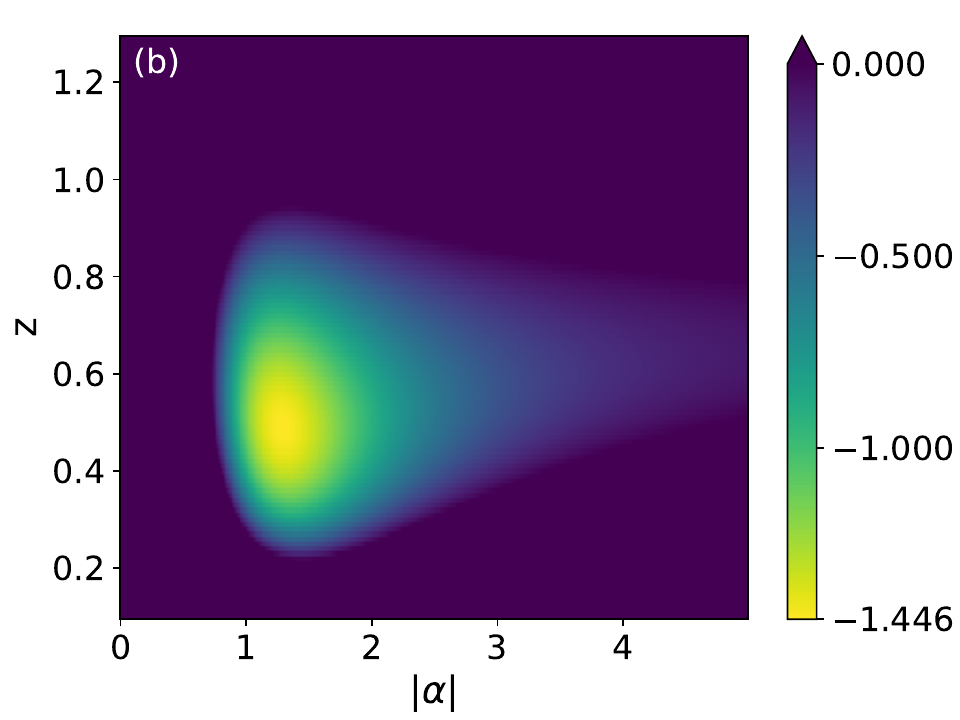}
\caption{(a) Wigner function of the PACS with the amplitude $\alpha = 1.43 + 0.65i$. The arrow $p^\prime$ shows the symmetry axis of the state, the tilt of which is determined by the coherent state amplitude. 
(b) Nonlinear squeezing \eqref{eq_nlsqdef} of $\ket{\phi(\alpha)}$ state shown in the dB scale, depending on the amplitude $|\alpha|$ and the cubicity $z$.}
\label{state_illustrations}
\end{figure}

A state that results in a variance of the quadrature operator being lower than the variance of the vacuum fluctuations is called a Gaussian squeezed state~\cite{Andersen2016}. Similarly, a state $\hat{\rho}$ that results in a variance of the operator~$\hat{O}$ being lower than the minimal variance over all Gaussian states is called a nonlinearly squeezed state. With this in mind, we can formally define the nonlinear squeezing by~\cite{Kala22,Brauer2021,Konno2021} 
\begin{equation}
\label{eq_nlsqdef}
 \xi_{\hat{\rho}} = \min_{z, \theta}  \Bigg[ \frac{\mathrm{var}_{\hat{\rho}}  (\hat{O}(z,\theta))}{\min_{\mathcal{G}}[\mathrm{var}_{\mathcal{G}}(\hat{O}(z,\theta))]} \Bigg],
\end{equation}
where the minimization over the angle $\theta$ is equivalent to the rotation of the state in the phase space, see Fig.~\ref{state_illustrations}(a).
The denominator corresponds to the minimum achievable over the set of all Gaussian states which is $\min_{\mathcal{G}}[\mathrm{var}_\mathcal{G}(\hat{O}(z, \theta))] = 3\big(\frac{1}{2} \big)^{ \frac{5}{3}} |z|^{\frac{2}{3}}$~\cite{Kala22}. 

The simplest state, for which the nonlinear squeezing can be observed, is a coherent superposition of the vacuum $\ket{0}$ and single photon $\ket{1}$, namely,
\begin{equation}
    \ket{\tilde{\phi}(c)} = \eta(c) \Big[  c\ket{0} +  \ket{1} \Big],
    \label{eq_01_single_mode}
\end{equation}
where $\eta(c)=(\sqrt{|c|^2 + 1})^{-1}$ is the normalization coefficient and $c$ is the complex amplitude~\cite{Miyata2016}.
A state similar to \eqref{eq_01_single_mode} can be generated using the quantum-optical catalysis technique~\cite{Lvovsky2002} or the photon added coherent state (PACS) technique, where a single photon is added to a coherent state $\ket{\alpha}$~\cite{Bellini2004,Tara1991,Fadrny2024} 
:
\begin{equation} 
    \ket{\Psi(\alpha)} = \eta(\alpha) \: \hat{a}^{\dagger}\ket{\alpha} =   \hat{D}(\alpha) \eta(\alpha)  \Big[  \alpha^* \ket{0} +  \ket{1} \Big] = \hat{D}(\alpha) \ket{\phi(\alpha)},
    \label{eq_pacs_single_mode}
\end{equation}
where  $\hat{D}(\alpha) = \exp(\alpha \hat{a}^{\dagger}-\alpha^*\hat{a})$ is the displacement operator, and $\eta(\alpha)$ again denotes the normalization term.

It should be noted that according to \eqref{eq_pacs_single_mode}, the tilt of the PACS symmetry axis, see Fig.~\ref{state_illustrations}(a), is determined by the amplitude of the coherent state $\alpha=|\alpha|e^{i \theta}$, the phase of which defines an angle $\theta=\theta_{min}$ that minimizes the variance of the nonlinear operator $\hat{O}$ in \eqref{eq_01_single_mode}. One can always rotate the quadrature phase space to align its $p$ axis with the symmetry axis $p^\prime$. Then the PACS becomes symmetric with respect to $x^\prime\rightarrow-x^\prime$ exchange.
Due to this symmetry, the displacement operator in \eqref{eq_pacs_single_mode} acting along the symmetry axis does not change the nonlinear squeezing measured (see Appendix~\ref{appendix_A}). As a result, $\xi_{\ket{\Psi(\alpha)}} = \xi_{ \ket{\phi(\alpha)} }$. 

Due to the basis truncation up to one photon, the value of nonlinear squeezing of $\ket{\phi(\alpha)}$  state is bounded by  $\xi_{\phi} = -1.45$~dB  (realized for $|\alpha| = 1.28$ and cubicity $z = 0.49$)~\cite{Miyata2016},  
see Fig.~\ref{state_illustrations}(b).
The experimental realization of PACS is based on the spontaneous parametric down-conversion (SPDC) process, in which
two generated photons appear in a set of broadband modes, known as the Schmidt modes~\cite{Sharapova2015,Sharapova2018}, that determine the spatio-temporal properties of the output light.

In this paper, we study the generation of states with nonlinear squeezing using multimode type-II SPDC and PACS technique. We investigate how the shape of the seed and local oscillator affects the magnitude of nonlinear squeezing. In the case of a two-mode source, we demonstrate how to create nonlinear squeezing in two modes simultaneously. For a real multimode source, we study the degree of the nonlinear squeezing with respect to the number of Schmidt modes of the system.

\begin{figure}[ht]
\centering
\includegraphics[width=0.99\linewidth]{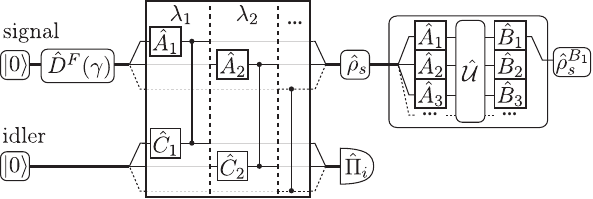}
\caption{The sketch of the studied process. The initial coherent state with an amplitude $\gamma$ in a broadband mode $\hat{F}$ seeds the PDC process in the signal channel. At the output, the mode-unresolved single-photon detection in the idler channel $\hat{\Pi}_i$ is performed yielding a photon addition into the signal channel. Here,  $\hat{A}_n$ and $\hat{C}_n$ are the Schmidt modes of the signal and idler photons, respectively, while $\lambda_n$ are the weights of the Schmidt decomposition.
To study the nonlinear squeezing in an arbitrary broadband mode $\hat{B}_1$ of the generated state $\hat{\rho}_s$, the linear transformation $\hat{\mathcal{U}}$ between the PDC Schmidt modes and measurement basis is performed resulting in the density matrix~$\hat{\rho}^{B_1}_s$. 
}
\label{fig_scheme}
\end{figure}

The scheme under the study is shown in Fig.~\ref{fig_scheme}. 
The unitary evolution operator of the PDC process yields $\hat{U}_{pdc} = \exp\big[ \Gamma \int d \omega_s d \omega_i S(\omega_s,\omega_i)\hat{a}^{\dagger} (\omega_s)\hat{c}^{\dagger} (\omega_i)+h.c.\big]$~\cite{Fabre2020,Roeland2022}, where
 $\hat{a}^{\dagger}(\omega_s)$ and $\hat{c}^{\dagger}(\omega_i)$ are the monochromatic creation operators of the signal and idler subsystems, respectively, and $\Gamma$ is the parametric gain.
In the low-gain regime ($\Gamma \ll 1$), one can expand the evolution operator up to the  first order and apply
the Schmidt decomposition to the joint spectral amplitude
$
  S(\omega_s,\omega_i)  =   \sum_n \sqrt{\lambda_n} \tau_n(\omega_s) \zeta_n(\omega_i),  
$
where  $\lambda_n$ are the eigenvalues obeying  $\sum_n\lambda_n=1$ and $\tau_n(\omega_s),  \zeta_n(\omega_i) $ are the eigenfunctions.
Then
the evolution operator can be written in a diagonalized form as
\begin{equation}
\label{eq_evolution_pdc}
\begin{split}
    \hat{U}_{\textrm{PDC}}  &\approx  \hat{I} + \Gamma \int d \omega_s d \omega_i S(\omega_s,\omega_i)\hat{a}^{\dagger} (\omega_s)\hat{c}^{\dagger} (\omega_i)+h.c. \\
    &\approx  \hat{I} + \Gamma \sum_n \sqrt{\lambda_n} \hat{A}^{\dagger}_n \hat{C}^{\dagger}_n   +h.c., 
\end{split}
\end{equation}
where $\hat{A}^{\dagger}_n = \int d\omega \tau_n(\omega)\hat{a}^{\dagger}(\omega)$ 
and $\hat{C}^{\dagger}_n = \int d\omega \zeta_n(\omega)\hat{c}^{\dagger}(\omega)$
are the new broadband Schmidt operators for the  signal and idler subsystems, respectively.   
An effective number of the occupied Schmidt modes is defined by the Schmidt number $K=1/\sum_n\lambda^2_n$.

A coherent broadband seed with a spectral profile $f(\omega)$ and an amplitude $\gamma$ in the signal channel $\ket{\gamma}^F=\hat{D}^F(\gamma)\ket{0}$  can be introduced via the displacement operator $\hat{D}^F(\gamma) = \exp(\gamma \hat{F}^{\dagger}-\gamma^*\hat{F})$ 
defined in terms of the broadband mode $\hat{F} = \int d\omega f(\omega) \hat{a}(\omega)$ with $\int d\omega |f(\omega)|^2 = 1$.  
This broadband mode can be decomposed with respect to the PDC  Schmidt-mode basis as $\hat{F} = \sum_m f^A_m \hat{A}_m$, where the expansion coefficients read $f_m^A=\int d\omega \tau^*_m(\omega) f(\omega)$. Then, the displacement operator has a form
\begin{equation}
    \hat{D}^F(\gamma) = \bigotimes_m \hat{D}^A_m(\alpha_m),
\end{equation}
where $\hat{D}^A_m(\alpha_m) = \exp \Big( \alpha_m  \hat{A}_m^{\dagger}-\alpha^*_m\hat{A}_m \Big) $ is the displacement operator in the broadband Schmidt mode of order $m$, while $\alpha_m = \gamma f^A_m$ is an  amplitude of the coherent state  in  mode $m$.

Therefore, the state generated in the PDC process with a broadband coherent seed can be fully represented in the basis of Schmidt modes as
\begin{equation} 
   |\psi \rangle  = \hat{U}_{\textrm{PDC}} \bigotimes_m \hat{D}_m(\alpha_m) \ket{0}. 
\end{equation}
To create the PACS, the heralding with the use of the  single-photon detector in the idler channel is performed.
In this work, we consider the simplest case of mode-unresolved single-photon detector described by the POVM
 $\hat{\Pi}_i =  \sum_k ~ \hat{C}_k^{\dagger}\dyad{0}\hat{C}_k$.
The resulting density matrix in a signal subsystem has the form~\cite{Roeland2022}  
\begin{equation}
   \rho_s = \frac{\mathrm{Tr}_i[ \hat{\Pi}_i \dyad{\psi} ]}{\mathrm{Tr}[ \hat{\Pi}_i \dyad{\psi} ]}  
   \label{eq_detection_state}
\end{equation}
and explicitly reads
\begin{equation}
    \hat{\rho}_s = 
    \mathcal{N} \sum_n   \bigotimes_m \lambda_n \hat{A}^{\dagger}_n  \hat{D}_m(\alpha_m) \ketbra{0} \hat{D}^\dagger_m(\alpha_m)  \hat{A}_n,
\label{eq_outputstate_schmidt}
\end{equation}
where 
$\mathcal{N} = \Big( \sum_n \lambda_n (|\alpha_n|^2+1)  \Big)^{-1} $ is the normalization coefficient and $ \bigotimes_{m,m^\prime} \ket{\alpha_m}\bra{\alpha_m^\prime}=\bigotimes_{m} \ketbra{\alpha_m}$ is taken into account.

The resulting multimode state has the following properties.
As opposed to the squeezed state (that describes the quadrature squeezing) which is pure and factorizable in the Schmidt-mode basis, the state \eqref{eq_outputstate_schmidt} is a statistical mixture of PACS in different Schmidt modes and is not factorizable.
This appears due to the mode unresolved-detection (\eqref{eq_detection_state}) used in the protocol.
The pure state in the form of \eqref{eq_pacs_single_mode} can only be generated in the case of the single-mode PDC (when only one Schmidt mode is populated $\lambda_n=\delta_{1n}$) seeded by the coherent light with a profile of PDC mode  $\alpha_m = \gamma \delta_{1m}$, where $\delta_{1m}$ is the Kronecker symbol.

Having a multimode scenario, the question of the nonlinear squeezing extraction  comes into focus.
The most straightforward solution to this problem seems to be the use of a broadband modes demultiplexing, namely, the splitting of each broadband mode to a set of independent optical channels with their own detectors~\cite{Serino2023,Brecht2015,Reddy2018}.
However, demultiplexing of a multimode field requires quantum pulse gates based on quantum frequency conversion, which is limited by the bandwidths of the interacting fields and introduces significant optical losses~\cite{Eckstein2011}.

In opposite to demultiplexing, where each mode is explicitly extracted from the initial multimode beam, the field quadratures of the chosen mode can be measured via homodyne detection applied to the full multimode beam.
Such type of detection not only enables state tomography and Wigner-function reconstruction~\cite{Jezek2003}, but is also an important tool for the direct nonlinear squeezing measurement~\cite{Kala22} and measurement-based cubic gates~\cite{Miyata2016}.
Homodyne detection effectively projects multimode light into the local oscillator mode $\hat{\mathcal{A}}_{LO}$, which is usually taken as the pump mode, but can be arbitrarily shaped using a spatial light modulator.

The measurement basis of the local oscillator $\hat{B}$ can be connected with the Schmidt-mode basis of PDC using the unitary matrix $U$: 
\begin{equation}\label{measurement_basis}
    \hat{A}_n = \sum_l  u_{nl} \hat{B}_l. 
\end{equation}
To present the state \eqref{eq_outputstate_schmidt} in the measurement basis $\hat{B}$, we rewrite the displacement operator in this basis as $\hat{D}^F(\gamma)= \bigotimes_m \exp \Big( \beta_m  \hat{B}_m^{\dagger}-\beta^*_m\hat{B}_m \Big)$, 
where the amplitudes $\beta_m =  \sum_n u^*_{nm} \alpha_n$. Then, the output state \eqref{eq_outputstate_schmidt} is given by
\begin{equation}
    \hat{\rho}_s = 
        \mathcal{N} \sum_{l l^\prime} \bigotimes_m   c_{ll^\prime} \hat{B}^\dagger_l  \hat{D}^B_m(\beta_m) \ketbra{0} [\hat{D}^B_m(\beta_m)]^\dagger   \hat{B}_{l^\prime}.
        \label{eq_outputstate_arbitrary}
\end{equation}
with coefficients  $c_{ll^\prime} = \sum_n \lambda_n u^{*}_{nl} u_{nl^\prime} $.

Assuming that the homodyne measurement is performed in only the $j$-th mode of the measurement basis, $\hat{\mathcal{A}}_{LO} = \hat{B}_j$, we can reduce the density matrix \eqref{eq_outputstate_schmidt} over the modes $\{ \hat{B}_1, ...\hat{B}_{j-1}, \hat{B}_{j+1}, ...  \}$ as the partial trace $\hat{\rho}^{B_j} \equiv \hat{\rho}^{LO} = \mathrm{Tr}_{\hat{B}_1, ...\hat{B}_{j-1}, \hat{B}_{j+1}, ... }(\rho_s)$.
The resulting state in the $j$-th mode is given by 
\begin{equation}
     \hat{\rho}^{B_j} \equiv \hat{\rho}^{LO} = \hat{D}^B_j(\beta_j) \hat{\rho}_{0j} [\hat{D}^B_j(\beta_j)]^\dagger, 
     \label{eq_state_LO}
 \end{equation} 
where
\begin{align}
    \label{stateinbasis}
    \hat{\rho}_{0j} = \mathcal{N} \Big[ & c_{jj} \ket{\phi(\beta_j)}\bra{\phi(\beta_j)}  + G_j \ket{\phi(\beta_j)}\bra{0} 
       \nonumber
     \\
   & + G_j^* \ket{0}\bra{\phi(\beta_j)}  + \eta_j \ket{0}\bra{0} \Big],  
\end{align}
while $\ket{\phi(\beta_j)} = \beta_j^*\ket{0} + \ket{1}$, $G_j =  \sum_{m\neq j} c_{jm}\beta_m $ and $\eta_j     = 1/\mathcal{N} - c_{jj} (|\beta_j|^2+1) - 2 \ \textrm{Re} (\beta_j^* G_j)$. Note that the arising nonlinear squeezing is associated with the state $\ket{\phi(\beta_j)}$, therefore, in the same way as for the pure state \eqref{eq_pacs_single_mode}, the displacement operator $\hat{D}^B_j(\beta_j)$ does not change the nonlinear squeezing value: $\xi_{\hat{\rho}_{s,j}^{LO}} = \xi_{  \hat{\rho}_{0j} }$. 
The state in the form of \eqref{eq_state_LO} allows us to compute the nonlinear squeezing in an arbitrary measured mode given by the local oscillator $\hat{\mathcal{A}}_{LO}$. As will be shown below, the shape of the local oscillator plays an important role in extracting maximum nonlinear squeezing from the system.

 \subsection{Single-mode case}

 In the case of a single-mode PDC, where only the Schmidt-mode $\hat{A}_1$ is populated $\lambda_n = \delta_{1n}$, the best value of nonlinear squeezing can be obtained by using the seed and the local oscillator in the same Schmidt mode: $\hat{F}=\hat{B}_1=\hat{A}_1$, since seeding and measurement in any other mode add noise to the system that reduces the nonlinear squeezing value. In this case, the density matrix \eqref{stateinbasis} describes the pure state with nonlinear squeezing depicted in Fig.~\ref{state_illustrations}.

 \subsection{Two-mode case}

Let us now consider the case, where only two modes $\hat{A}_1$ and $\hat{A}_2$  with eigenvalues $\lambda_1$ and $\lambda_2=1- \lambda_1  $, respectively,  are populated. A two-mode coherent seed can be represented in a general form as 
 $\hat{F} = f_1\hat{A}_1+f_2\hat{A}_2 $ with $|f_1|^2 + |f_2|^2 = 1$. For simplicity the amplitudes $f_1$ and $f_2$ are taken to be real and $f_2=\sqrt{1-f_1^2}$.

According to \eqref{measurement_basis}, the two-mode measurement basis  $\hat{B}_l=   \sum_n  u^*_{nl} \hat{A}_n $ can be parameterized using the rotating angle $\nu$ as 
\begin{equation}
\begin{split}
    \hat{B}_1= \mathrm{cos}[\nu] \hat{A}_1 + \mathrm{sin}[\nu] \hat{A}_2 
    \\
   \hat{B}_2= \mathrm{cos}[\nu] \hat{A}_2 -  \mathrm{sin}[\nu] \hat{A}_1.
   \end{split}
   \label{basis_B}
\end{equation}

\begin{figure}[ht]
\includegraphics[width=0.49\linewidth]{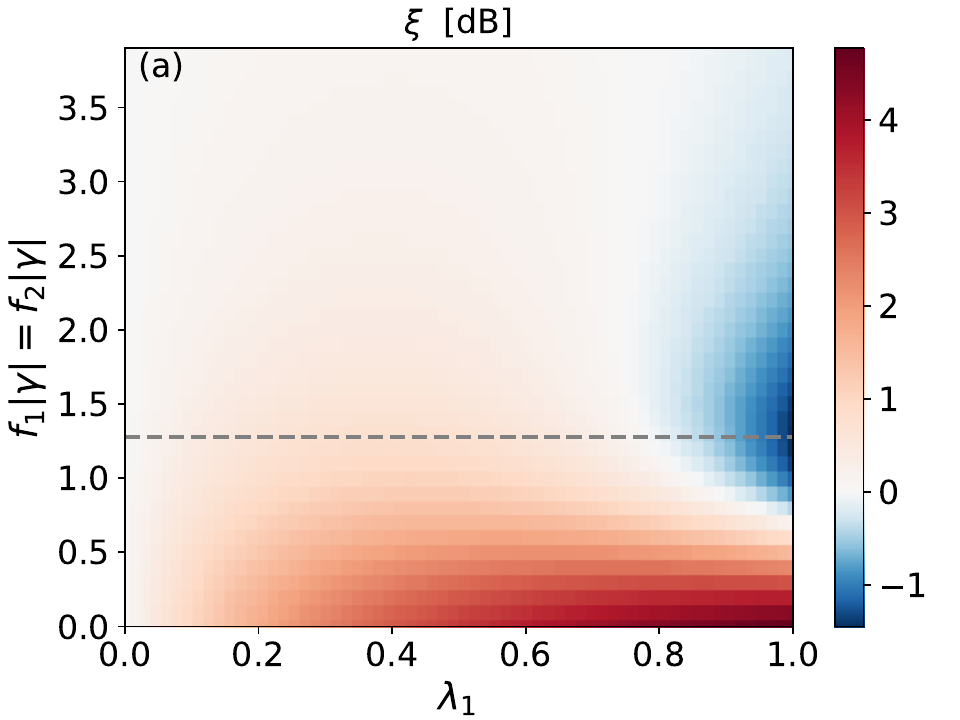}
\includegraphics[width=0.49\linewidth]{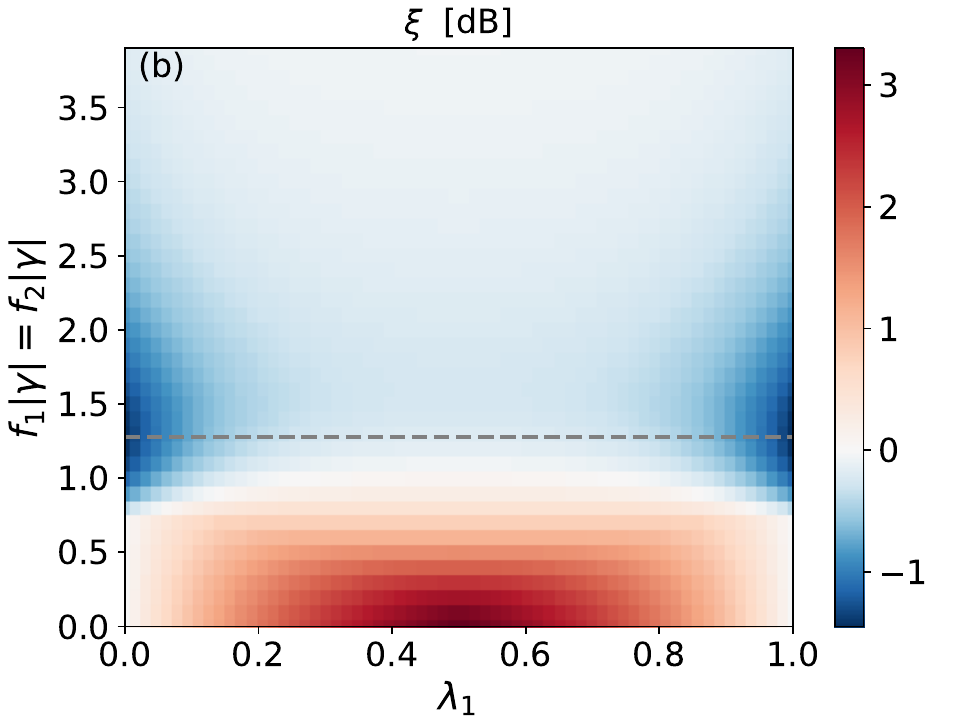}
\includegraphics[width=0.49\linewidth]{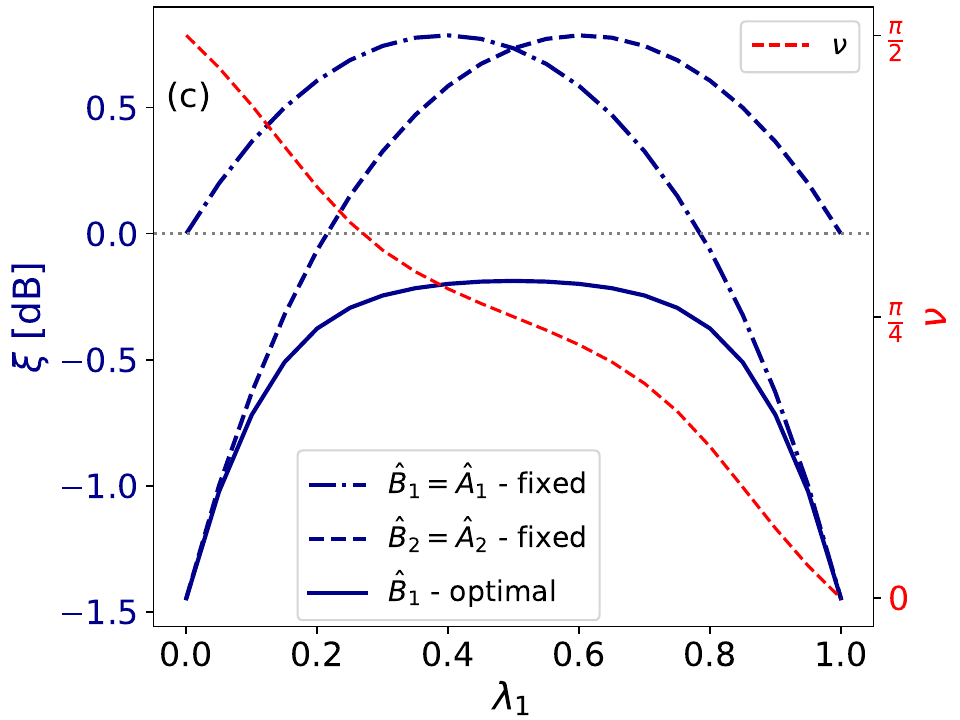}
\includegraphics[width=0.49\linewidth]{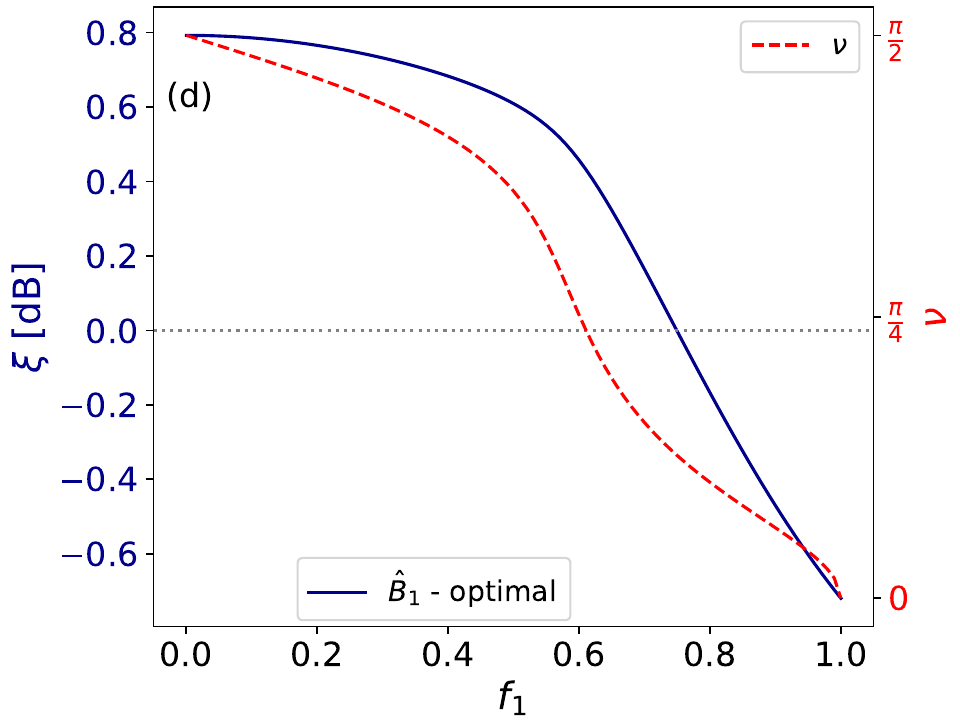}
\caption{Nonlinear squeezing in the first mode $\hat{B}_1$ of the two-mode PACS.
 (a) $\hat{B}_1=\hat{A}_1$ , (b) $\hat{B}_1$ is optimized by rotating $\nu$ in \eqref{basis_B};
 (c) cross-section of (a) (blue dashed-dotted line) and (b) (blue solid line) along the marked dashed gray line $|\alpha_1|=|\alpha_2|= |\gamma|/\sqrt{2}=1.28$. The dashed blue line depicts the nonlinear squeezing in the second mode $\hat{B}_2=\hat{A}_2$.  The red dashed line, tracked on the right vertical axis, shows the optimal rotation angle $\nu$. 
 (d) Nonlinear squeezing in the optimized mode $\hat{B}_1$ and the optimal angle $\nu$ versus the weight parameter of the seed, $|\gamma| = 1.28 $, $\lambda=0.8$. For all plots the cubicity is optimized individually for each point.}
\label{fig_two_modes}
\end{figure}

If $\nu=0$, the measurement basis coincides with the Schmidt-mode basis, $\hat{B}_n=\hat{A}_n$, while all coefficients in \eqref{stateinbasis} are significantly simplified and are given by $c_{11}=\lambda_1$, $c_{22}=\lambda_2$, $G_1=G_2=0$. Then \eqref{stateinbasis} presents an incoherent mixture of the PACS $\ket{\phi(\beta_j)}$ and vacuum:

\begin{equation}\label{eq_twomodes_dm}
    \hat{\rho}_{0j} = \mathcal{N} \Big[ \lambda_j \ket{\phi(\beta_j)}\bra{\phi(\beta_j)}  +  (1-\lambda_j) \big((1-|f_j|^2)|\gamma|^2+1\big)\ket{0}\bra{0} \Big].  
\end{equation}

Considering a seed which occupies both modes equally, $f_1=f_2=1/\sqrt{2}$, which leads to  $\alpha_1=\alpha_2= \gamma/\sqrt{2}$, the state in \eqref{eq_twomodes_dm} for $j=1$ (measurement in $\hat{B}_1$)
results in the nonlinear squeezing presented in Fig.~\ref{fig_two_modes}(a). 
One can notice that as soon as the population $\lambda_1$ does not become equal to one, the state \eqref{eq_twomodes_dm} ceases to be pure and the nonlinear squeezing in the first mode decreases.  
For the population below approximately $\lambda_1=0.8$, we are unable to unveil any non-Gaussianity. The measurement in the second mode $\hat{B}_2$ ($j=2$ in \eqref{eq_twomodes_dm}) reveals a map of  nonlinear squeezing inverted in  $\lambda_1$  with respect to $\hat{B}_1$, because the density matrix in \eqref{eq_twomodes_dm} obeys the symmetry $ \hat{\rho}_{01}(\lambda_1)=\hat{\rho}_{02}(1-\lambda_1)$.
The blue dashed-dotted line in Fig.~\ref{fig_two_modes}(c) presents a cut of Fig.~\ref{fig_two_modes}(a) along the marked line with $|\gamma|=1.28 $. The dashed line of the same color presents the same cut, but for the measurement in $\hat{B}_2$.

For the same equally populated seed, the nonlinear squeezing measured in the first mode can be improved by optimizing the measurement basis, namely the parameter $\nu$, as illustrated  
 in Fig.~\ref{fig_two_modes}(b).
In this case, the nonlinear squeezing can be found for any population $\lambda_1$, showing 
that for a complicated seed profile, the Schmidt-mode basis is no longer an optimal basis for measuring the nonlinear squeezing.
The blue solid line in Fig.~\ref{fig_two_modes}(c) presents the cut of Fig.~\ref{fig_two_modes}(b) along the marked line with $|\gamma|=1.28$, while
 the red line shows the optimized values of  $\nu$ that were used to calculate this cut.
Since the measurement basis was optimized to obtain maximum squeezing in the first mode $\hat{B}_1$, the measurement in the orthogonal mode $\hat{B}_2$ does not show any nonlinear squeezing and is therefore not plotted.

To investigate the dependence of the nonlinear squeezing on the seed profile, we fix the mode population $\lambda_1=0.8$ and the seed amplitude  $|\gamma| = 1.28$. Fig.~\ref{fig_two_modes}(d) shows the nonlinear squeezing measured in  $\hat{B}_1$ as a function of the population amplitude of the first seed mode $f_1$, while the cubicity $z$ and the measurement basis $\nu$ are optimized. 
It can be seen that to obtain the nonlinear squeezing, the seed amplitude in the first mode $\alpha_1 = \gamma f_1 $ must be high enough. If we reduce the population amplitude $f_1$ for a fixed coherent state amplitude $\gamma$, the coherent seed amplitude in the first mode becomes insufficient  to generate any nonlinear squeezing, even optimizing the measurement mode.

\begin{figure}[ht]
\includegraphics[width=0.49\linewidth]{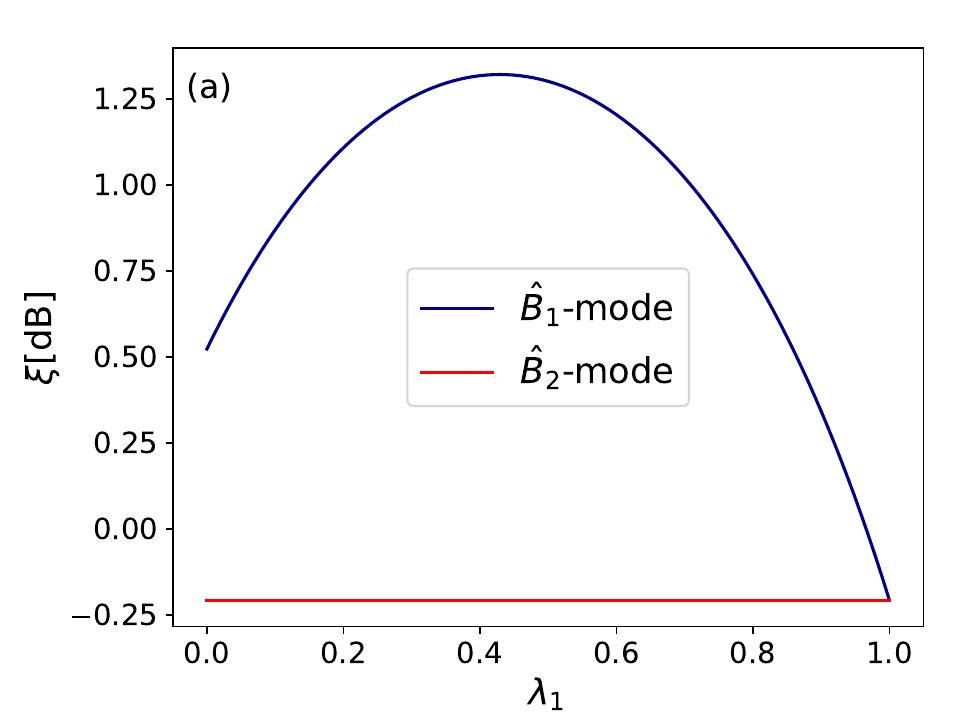}
\includegraphics[width=0.49\linewidth]{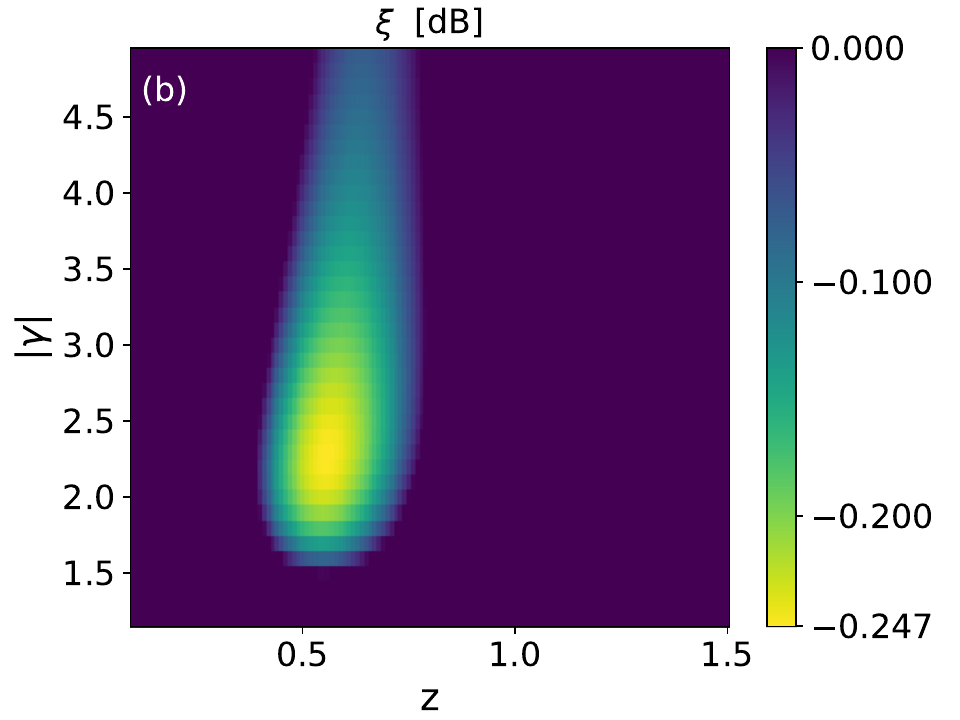}
\caption{a) Nonlinear squeezing in two modes simultaneously, the measurement basis is rotated by $\nu=-\frac{\pi}{4}$, the seed occupies both modes equally $f_1=f_2 = \frac{1}{\sqrt{2}}$.
b)  Nonlinear squeezing for the point $\lambda_1=1$ in (a), where the squeezing in the first and second modes coincides, versus $\gamma$ and $z$.
}
\label{two_mode}
\end{figure}

Finally, we study the possibility of generating nonlinear squeezing in both modes simultaneously. For this, we consider the same seed as for Fig.~\ref{fig_two_modes}(a-c), namely  $\alpha_1=\alpha_2= \gamma/\sqrt{2}$, but a specific basis rotated by $\nu=-\frac{\pi}{4}$.
In such a basis, the coherent seed amplitude of the first measurement mode $\beta_1= 0$, while for the second mode we get $\beta_2= \gamma$. This results in the off-diagonal coefficients of the form  $G_1=\frac{\lambda_1-\lambda_2}{2 }\gamma$ and $G_2=0$, while  the coefficients  $c_{11}=c_{22}=\eta_2=1/2$ do not depend on the populations $\lambda$. 
Thus, one can explicitly write the reduced density matrix \eqref{eq_twomodes_dm} for the first

\begin{equation}\label{eq_twomodes_dm_1}
    \hat{\rho}_{01} = \frac{ \Big[ \frac{1}{2} \ket{1}\bra{1}  + \frac{\lambda_1-\lambda_2}{2}\gamma\ket{1}\bra{0} + \frac{\lambda_1-\lambda_2}{2 }\gamma^*\ket{0}\bra{1} +  \frac{|\gamma|^2+1}{2} \ket{0}\bra{0} \Big]}{ \frac{|\gamma|^2}{2}+1} 
\end{equation}
and the second 
\begin{equation}\label{eq_twomodes_dm_2}
    \hat{\rho}_{02} = \frac{ \Big[ \frac{1}{2} \ket{1}\bra{1}  + \frac{1}{2}\gamma\ket{1}\bra{0} + \frac{1}{2 }\gamma^*\ket{0}\bra{1} +  \frac{|\gamma|^2+1}{2} \ket{0}\bra{0} \Big]}{ \frac{|\gamma|^2}{2}+1}.  
\end{equation}
measurement modes.
The associated with them nonlinear squeezing for the fixed cubicity $z=0.5$ and coherent seed amplitude $|\gamma|=2.5$ is presented in Fig.~\ref{two_mode}(a).

It can be seen that the matrix \eqref{eq_twomodes_dm_2} does not depend on the population coefficients  $\lambda$, therefore, as shown in  Fig.~\ref{two_mode}(a), the nonlinear squeezing in this mode is constant for all population weights. 
For $\lambda_1=1$, the reduced density matrices of both modes coincide $ \hat{\rho}_{01} =  \hat{\rho}_{02}$, see Fig.~\ref{two_mode}(a), and give the maximal amount of nonlinear squeezing that can be generated in both modes simultaneously.  A two-dimensional map of nonlinear squeezing versus the cubicity and coherent state amplitude for the fixed population $\lambda_1=1$ is shown in Fig.~\ref{two_mode}(b) and results in the maximal squeezing of -0.25 dB. 
In this case, having a single-mode PDC, we perform a measurement in the diagonal basis with respect to the Schmidt-mode basis, which results in a distribution of nonlinear squeezing between two modes.

\begin{figure}[ht]
\includegraphics[width=0.49\linewidth]{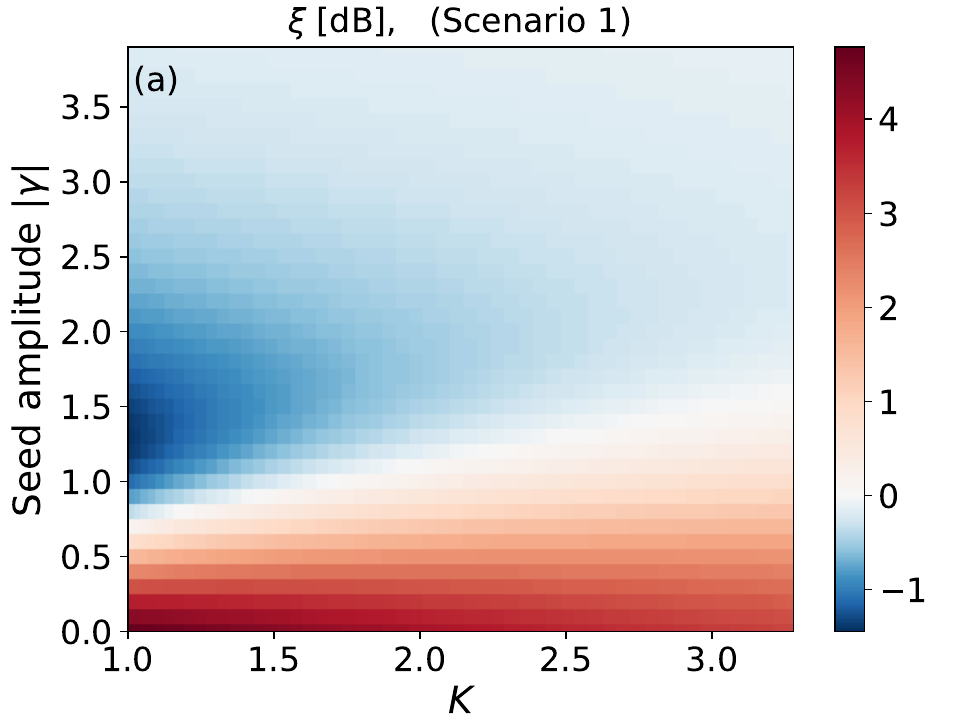}
\includegraphics[width=0.49\linewidth]{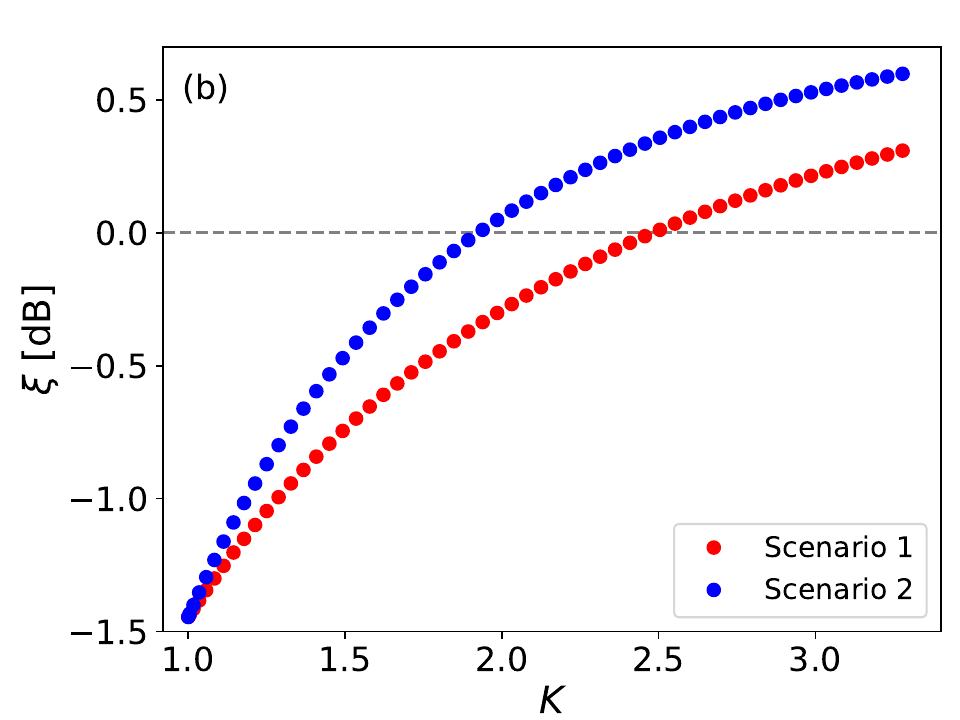}

\caption{
Nonlinear squeezing as a function of the number of Schmidt modes $K$. 
(a) The coherent seed and the local oscillator equal to the first Schmidt-mode of the multimode PDC source (Scenario 1).
(b) Cross-section of (a) at $|\gamma|=1.28 $ together with the
same cross-section for Scenario 2. 
}
\label{fig_multimode}
\end{figure}

\subsection{Multimode case}\label{multimodecase}

In order to study the nonlinear squeezing in a more realistic multimode scenario, we consider the joint spectral amplitude under the double-Gaussian approximation 
\begin{equation}
	S(\omega, \omega^\prime) = \exp(-\frac{(\omega+\omega^\prime)^2}{a^2})\exp(-\frac{(\omega-\omega^\prime)^2}{b^2}),
\end{equation}
where $a$ and $b$ are the spectral widths of the pump and phase-matching function, respectively.

In this case, the Schmidt number is defined as~\cite{Fedorov2009}

\begin{equation}
    K = \frac{1}{2} \Big( \dfrac{a}{b} + \dfrac{b}{a} \Big).
\end{equation}

In order to reduce an amount of parameters, we assume that the PDC source is fixed, i.e. the parameters $a$ and $b$ are fixed and therefore the number of modes $K$ is strictly defined.
Below we consider two experimental scenarios to get nonlinear squeezing in the first mode:
\begin{enumerate}
       \item Scenario 1: the coherent seed and the local oscillator are equal to each other and are simultaneously vary to coincide with the first Schmidt-mode $\hat{A}_1(a,b)$ of the multimode PDC source for each fixed parameters  $a$ and $b$, namely  $\hat{F}=\hat{B}_1=\hat{A}_1(a,b)$.
       \item Scenario 2: the coherent seed and the local oscillator are fixed and have a Gaussian profile of the Schmidt-mode of a single-mode PDC source, namely $\hat{F}=\hat{B}_1=\hat{A}_1(a=b)$.
\end{enumerate}

As demonstrated in Appendix~\ref{appendix_B}, the first scenario yields the optimal amount of nonlinear squeezing. For this case, the dependence of nonlinear squeezing on the number of Schmidt modes $K$ and the amplitude of coherent seed is presented in Fig.~\ref{fig_multimode}(a). As expected, in the case of $K=1$, the squeezing is maximized and equals to -1.45 dB.
By increasing the number of Schmidt modes $K$, the nonlinear squeezing decreases, however, it is still present for the multimode PDC regime.

The experimental realization of Scenario 2 is simpler compared to Scenario 1: In this case, the original unmodified laser field can be used for both the pump and seed light.  
Fig.~\ref{fig_multimode}(b) shows the cross-section of (a) at $|\gamma|=1.28 $ together with the same cross-section for  Scenario 2. 
The nonlinear squeezing for Scenario 2 is only slightly worse compared to the Scenario 1:
The nonlinear squeezing better then $-0.5$~dB can be observed for $K<1.4$ and $K<1.7$ for Scenarios 2 and 1, respectively.

\subsection{Direct homodyne measurement} 
Finally, we estimate the number of homodyne measurements required to detect the nonlinear squeezing.  Due to the symmetry properties of nonlinear squeezing, it is not necessary to perform the full tomography for all rotation angles: The variance of the operator $\hat{O}$ in \eqref{eq_nonlop} can be estimated from homodyne measurements of generazlized quadratures $X^\prime(\theta) = \cos(\theta)x^\prime + \sin(\theta)p^\prime$  at only four angles $\theta = (0, \pi/2, \pi/4,-\pi/4)$~\cite{Moore2019} of the local oscillator via 
\begin{equation}\label{var_4angles}
\begin{split}
\myvar_{\hat{\rho}} (\hat{p}^\prime + z \hat{x}^{\prime2}) = \expval{\hat{X}^\prime(\pi/2)^2} + z^2 \expval{\hat{X}^\prime(0)^4} \\+ \frac{2\sqrt{2}z}{3}[\expval{\hat{X}^\prime(\pi/4)^3}-\expval{\hat{X}^\prime(-\pi/4)^3}] \\
- \frac{2z}{3}\expval{\hat{X}^\prime(\pi/2)^3}- [\expval{\hat{X}^\prime(\pi/2)} + z \expval{\hat{X}^\prime(0)^2}]^2,
\end{split}
\end{equation}
where the quadrature $\hat{X}^{\prime}(0)=x^\prime$  minimizes the homodyne measurement output, while the quadrature $\hat{X}^{\prime}(\frac{\pi}{2})=p^\prime$  maximizes it.
In this case, the PACS shares the symmetry of operator $\hat{O}$ with respect to the  $x^\prime\rightarrow -x^\prime$  transformation. 

\begin{figure}[ht]
\includegraphics[width=0.49\linewidth]{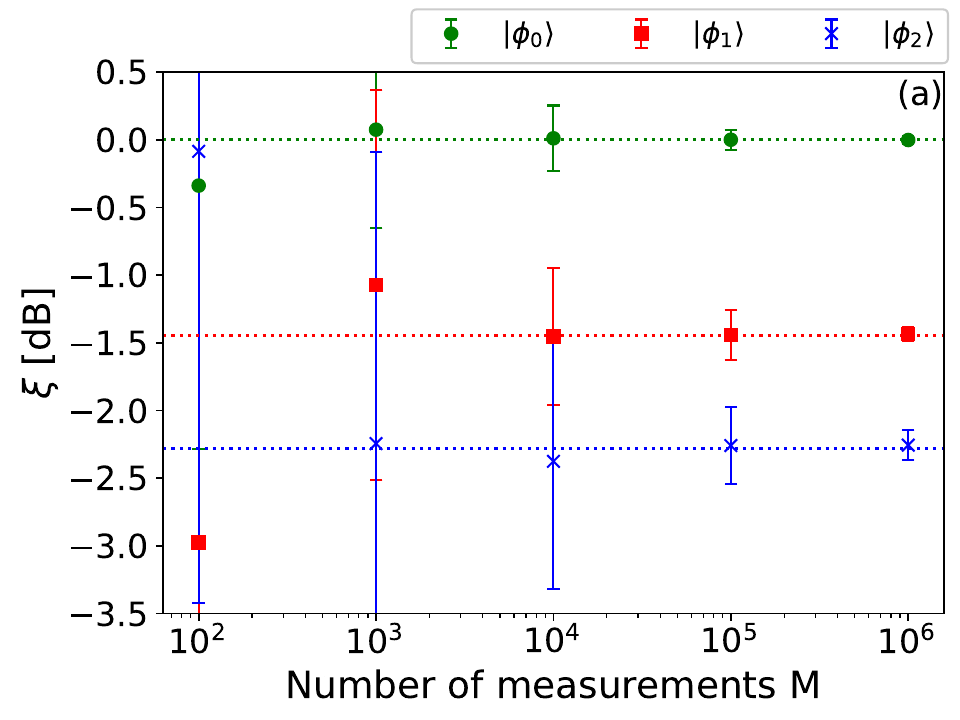}
\includegraphics[width=0.49\linewidth]{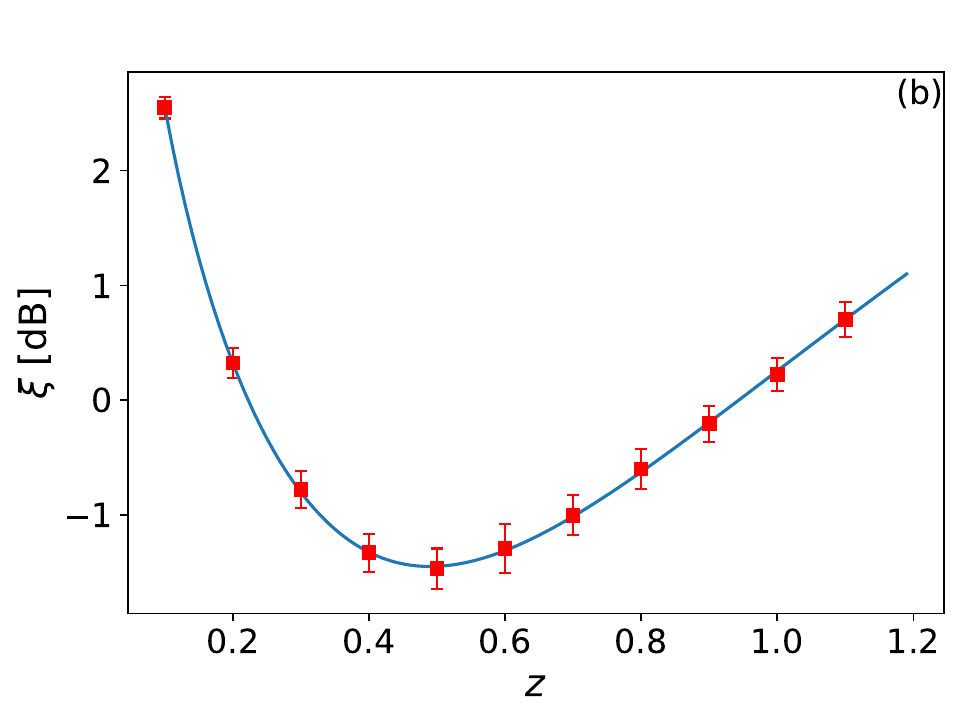}

\caption{ (a) Mean nonlinear squeezing estimated with the use of $N=100$ Monte Carlo simulations of \eqref{eq_nlsqdef}  together with its standard deviation for three different quantum states (see description in the text).
 Theoretically obtained values are shown by the dashed lines of the same color.
(b) Nonlinear squeezing obtained by the same Monte Carlo simulation with $M=10^5$ simulated measurements computed for different cubicities $z$. The blue line depicts theoretical values obtained by using the state \eqref{eq_01_single_mode} with $|c|=1.28$ in \eqref{eq_nlsqdef}. }
\label{fig_measn}
\end{figure}

The impact of limited number of measurements on the estimation of nonlinear squeezing from \eqref{var_4angles} is shown in Fig.~\ref{fig_measn}(a). Since the nonlinear squeezing grows with the number of Fock states in an optimized superposition~\cite{Miyata2016}, for our analysis we consider three states: 
a) the  vacuum state  $\ket{\phi_0}=\ket{0}$  (green)  that does not provide any nonlinear squeezing and therefore can serve as a check for false positive results of observing the non-Gaussianity; 
b)  the state of interest, namely \eqref{eq_01_single_mode},  $\ket{\phi_1} = \eta(c) \Big[  c\ket{0} +  \ket{1} \Big]$ (red);
c) the next state in the hierarchy $\ket{\phi_2} = \eta(c_1, c_2, c_3) \Big[  c_1\ket{0} +  c_2\ket{1} +c_3\ket{2} \Big]$ (blue), where $ \eta(c_1, c_2,c_3)= (\sqrt{|c_1|^2+|c_2|^2 + |c_3|^2})^{-1}$ is the normalization constant. 
The amplitudes of the last two states and the cubicity are optimized to get the highest nonlinear squeezing. As expected, the larger the number of photons in the superposition is, the higher the nonlinear squeezing becomes.

To calculate Fig.~\ref{fig_measn}, the analytical probability distributions of homodyne measurement for the four rotated quadratures were computed using the chosen wave function  as  $P_n(X(\theta)) = |\bra{x}\hat{R}(\theta)\ket{\phi_n}|^2$, where $\hat{R}(\theta)$ is the rotation in phase space by the angle $\theta$ and $n=0,1,2$. From these analytical distributions, $M$ values are randomly sampled to simulate $M$ homodyne measurement outcomes for a given state $\ket{\phi_n}$. The values are distributed in 1000 equal-width bins in an interval given by the smallest and largest generated values in the sample.  Within each bin the count $C$ (the number of sampling points inside the bin) is computed together with the mean value of the measurement outcome inside the bin
(using the function binned\textunderscore statistics from SciPy package). The mean values in each bin and the number of counts $C$ then approximate the initial probability distribution, however with limited accuracy determined by the number of measurements. From this approximate probability distribution, the moments appearing in \eqref{var_4angles} are computed. 
This procedure of randomly sampling $M$ values was repeated $N=100$ times
resulting in the mean nonlinear squeezing and its standard deviation  plotted in Fig.~\ref{fig_measn}(a). As the number of homodyne measurements for each rotated quadrature increases, the mean value approaches its theoretical value depicted by the dashed lines of the same color. Providing $M=10^5$ measurements for each of four homodyne angles, it is possible to distinguish the three states within the standard deviation. 

Fig.~\ref{fig_measn}(b) shows a dependence of the estimated nonlinear squeezing on cubicity for the state \eqref{eq_01_single_mode} and $M=10^5$ measurements for each local oscillator angle (red dotes) together with its analytical evaluation (blue line) for the amplitude $|c|=1.28$ and \eqref{eq_nlsqdef}. The minimum is obtained for the states native cubicity~\cite{Kala2024}. 

In conclusion, we described the properties and behaviour of nonlinear squeezing in a multimode scenario. In particular,  we analyzed the nonlinear squeezing of states created by the multimode single-photon addition with the use of seed multimode PDC. In a two-mode case, we demonstrated regimes in which the  nonlinear squeezing can be detected in both modes simultaneously and non-Gaussian properties are spread among the modes. Furthermore, we showed the importance of the choice of measurement modes and the possibility of its optimization in order to maximize the nonlinear squeezing detected. 
Finally, we discussed a  realistic setup with a multimode PDC source using the double-Gaussian approximation of the joint spectral amplitude. Here, the nonlinear squeezing generally decreases with the Schmidt number. However, up to $K=1.7$ a reasonable amount of nonlinear squeezing can be detected in at least one mode. We conclude by discussion of the nonlinear squeezing detection via homodyne measurement technique.

\begin{backmatter}
\bmsection{Funding}
D.K. and P.S. acknowledge the support by the `Photonic Quantum Computing' (PhoQC) project, which is funded by the Ministry for Culture and Science of the State of North-Rhine Westphalia.
P.S. acknowledges financial support of the Deutsche Forschungsgemeinschaft (DFG) via Project SH 1228/3-1 and via the TRR 142/3 (Project No. 231447078, Sub- project No. C10). We also thank the PC2 (Paderborn Center for Parallel Computing) for providing computation time.
{P.M. and V.K acknowledge the support by the project  Grant No. 22-08772S of the Czech
Science Foundation, the European Union’s HORIZON
Research and Innovation Actions under Grant Agreement
no. 101080173 (CLUSTEC). P.M. acknowledges the project (QUEENTEC); OP JAK CZ.02.01.01 \slash 00 \slash 22\textunderscore008 \slash 0004649 funded by Ministry of Education, Youth and Sport of the Czech Republic and V.K. acknowledges the project IGA-PrF-2024-008}





\end{backmatter}

\bibliography{references} 
\newpage
\appendix

\section{Independence of the nonlinear squeezing on displacement}
\label{appendix_A}

Considering the state $\hat{\rho}$ whose Wigner function is symmetric under the transformation $x \rightarrow -x$ , the angle $\theta$ in \eqref{eq_nlsqdef} can be omitted and the variance of the operator $\hat{O}$ can be rewritten as
\begin{equation}\label{rewritten}
    \textrm{var}(\hat{p} + z\hat{x}^2)=\myvar(\hat{p}) + 2z \textrm{cov}(\hat{p},\hat{x}^2) + z^2\myvar(\hat{x}^2).
\end{equation}
The displacement in $p$ transforms the $\hat{p}$ quadrature as $\hat{p}\rightarrow \hat{p} + d_p$, where $d_p$ is an amount of displacement, leaving the $\hat{x}$ quadrature unchanged. Therefore the displacement in $p$ affects only the first two terms in the equation \eqref{rewritten}. The variance of $\hat{p}$ transforms as
\begin{equation}
\begin{split}
        &\textrm{var}(\hat{p}+d_p) = \langle(\hat{p}+d_p)^2\rangle-\expval{(\hat{p}+d_p)}^2\\
        &=\langle\hat{p}^2\rangle + 2d_p\expval{\hat{p}} + d_p^2 - \expval{\hat{p}}^2 - 2d_p\expval{\hat{p}} - d_p^2\\
        &=\textrm{var}(\hat{p}),
\end{split}
\end{equation}
where we take into account $\expval{d_p} = d_p$. The second term, covariance, changes as
\begin{equation}
    \begin{split}
    &\textrm{cov}(\hat{p}+d_p,\hat{x}^2)=\frac{1}{2}\langle(\hat{p}+d_p)\hat{x}^2+\hat{x}^2(\hat{p}+d_p)\rangle-\expval{\hat{p}+d_p}\langle\hat{x}^2\rangle\\
    &=\frac{1}{2}\langle\hat{p}\hat{x}^2+\hat{x}^2\hat{p}\rangle + d_p\langle\hat{x}^2\rangle-\expval{\hat{p}}\langle\hat{x}^2\rangle - d_p\langle\hat{x}^2\rangle\\
    &=\textrm{cov}(\hat{p},\hat{x}^2).
    \end{split}
\end{equation}
Therefore, the displacement along $p$-axis does not modify nonlinear squeezing. 

When the state is rotated in phase space, the same result holds for the displacement along $p^{\prime}$ (see main text and Fig. \ref{state_illustrations}).

\section{ Optimal nonlinear squeezing }
\label{appendix_B}
The explicit expression for the variance in the $j$-th mode reads
\begin{equation}
\begin{split}
 \mathrm{var}_{\hat{\rho}}  (\hat{O}_j(z,\theta)) = \Big(\frac{1}{2} + \mathcal{N} c_{jj}\Big) + z \frac{\mathcal{N}}{\sqrt{2}i}E_j
 +z^2\Big(\frac{3}{4}+3 \mathcal{N} c_{jj}\Big)
\\
-\Big(\frac{\mathcal{N}}{\sqrt{2}i}E_j+z\Big(\frac{1}{2} + \mathcal{N} c_{jj}\Big)\Big)^2, 
\label{Var}
 \end{split}
\end{equation}
where $E_j=2ic_{jj}\mathrm{Im}[\beta_j]+2i\mathrm{Im}[G_j] = 2i\sum_{n=1}^{\infty}  \mathrm{Im}[c_{jn} \beta_n]$.
Using the definition of amplitudes $\beta_n$ and coefficients $c_{jn}$, we get

\begin{equation}
\begin{split}
E_j=
2i\sum_{n=1}^{\infty}  \mathrm{Im}\Big[ \sum_k \lambda_k u^{*}_{kj} u_{kn} \sum_l u^*_{ln} \gamma f^A_l\Big]. 
\end{split}
\end{equation}
Taking into account that $\sum_{n} u_{kn} u^*_{ln} = \delta_{kl}$, where $\delta_{kl}$ is the Kronecker symbol, we obtain

\begin{equation}
\begin{split}
E_j=2i  \mathrm{Im}\Big[ \sum_k \lambda_k u^{*}_{kj}  \gamma f^A_k\Big].
\end{split}
\end{equation} 

Denoting $\kappa_j = \mathcal{N} c_{jj} $ and $\mu_j= \frac{\mathcal{N}}{\sqrt{2}i}E_j$, one can get 
\begin{equation}
 \mathrm{var}_{\hat{\rho}}  (\hat{O}_j(z,\theta)) =  -\mu^2_j -  \mu_j (2 z \kappa_j ) + \Big(\frac{1}{2} + \kappa_j 
 +\frac{z^2}{2}+2z^2 \kappa_j  -z^2 \kappa_j^2\Big),
\label{Var1}
\end{equation}
which means that the variance has an  inverse parabolic dependence with respect to the parameter $\mu_j$. For such a dependence, the variance is minimized when $\mu_j$ is maximized. Considering the explicit form $\mu_j=\frac{\sqrt{2}\mathrm{Im}\Big[ \sum_k \lambda_k u^{*}_{kj}  \gamma f^A_k\Big]}{1+ \sum_n \lambda_n  |f^A_n\gamma|^2 }$, one can notice that for the fixed coherent seed amplitude $\gamma$ and Schmidt weights $\lambda_n$,  the parameter $\mu_j$ is maximized when $u_{kj}$ and $f^A_j$ reach their maximal value, namely $u_{kj}=\delta_{1j}$ and $f^A_j=1$, which means that the measurement in the Schmidt-mode basis of the crystal together with the seed in the chosen Schmidt mode $\hat{A}_j$ result in the optimal nonlinear squeezing in mode $\hat{A}_j$.

\end{document}